\documentclass[preprint]{emulateapj}

\received{}
\revised{}
\accepted{}

\shortauthors{Boyer et al.}
\shorttitle{``Dust Production in NGC 362''}

\begin{document}

\title{Dust Production and Mass Loss in the Galactic Globular Cluster
NGC~362}

\author{Martha~L.~Boyer\altaffilmark{1},
  Iain~McDonald\altaffilmark{2}, 
  Jacco~Th.~van~Loon\altaffilmark{2},
  Karl~D.~Gordon\altaffilmark{1}, 
  Brian~Babler\altaffilmark{3},
  Miwa~Block\altaffilmark{4}, 
  Steve~Bracker\altaffilmark{3},
  Charles~Engelbracht\altaffilmark{4}, 
  Joe~Hora\altaffilmark{5},
  Remy~Indebetouw\altaffilmark{6}, 
  Marilyn~Meade\altaffilmark{3},
  Margaret~Meixner\altaffilmark{1}, 
  Karl~Misselt\altaffilmark{4},
  Joana~M.~Oliveira\altaffilmark{2},
  Marta~Sewilo\altaffilmark{1}, 
  Bernie~Shiao\altaffilmark{1}, and
  Barbara~Whitney\altaffilmark{1}}
  \altaffiltext{1}{STScI, 3700 San Martin Drive, Baltimore, MD 21218 USA; mboyer@stsci.edu}
  \altaffiltext{2}{Astrophysics Group, Lennard-Jones Laboratories, Keele University, Staffordshire ST5 5BG, UK}
  \altaffiltext{3}{Department of Astronomy, University of Wisconsin, Madison, 475 North Charter Street, Madison, WI 53706-1582 USA}
  \altaffiltext{4}{Steward Observatory, University of Arizona, 933 North Cherry Avenue, Tucson, AZ 85721 USA}
  \altaffiltext{5}{Harvard-Smithsonian Center for Astrophysics, 60 Garden Street, MS 65, Cambridge, MA 02138-1516 USA}
  \altaffiltext{6}{Department of Astronomy, University of Virginia, P.O. Box 3818, Charlottesville, VA 22903-0818 USA}

\begin{abstract}
We investigate dust production and stellar mass loss in the Galactic
globular cluster NGC~362. Due to its close proximity to the Small
Magellanic Cloud (SMC), NGC~362 was imaged with the IRAC and MIPS
cameras onboard the {\it Spitzer Space Telescope} as part of the
Surveying the Agents of Galaxy Evolution (SAGE-SMC) {\it Spitzer}
Legacy program. We detect several cluster members near the tip of the
Red Giant Branch that exhibit infrared excesses indicative of
circumstellar dust and find that dust is not present in measurable
quantities in stars below the tip of the Red Giant Branch. We modeled
the spectral energy distribution (SED) of the stars with the strongest
IR excess and find a total cluster dust mass-loss rate of
$3.0^{+2.0}_{-1.2} \times 10^{-9} M_\odot~{\rm yr^{-1}}$, corresponding
to a gas mass-loss rate of $8.6^{+5.6}_{-3.4} \times 10^{-6} M_\odot~{\rm
yr^{-1}}$, assuming [Fe/H] $= -1.16$. This mass loss is in addition to
any dust-less mass loss that is certainly occurring within the
cluster. The two most extreme stars, variables V2 and V16, contribute
up to 45\% of the total cluster dust-traced mass loss. The SEDs of the
more moderate stars indicate the presence of silicate dust, as
expected for low-mass, low-metallicity stars.  Surprisingly, the SED
shapes of the stars with the strongest mass-loss rates appear to
require the presence of amorphous carbon dust, possibly in combination
with silicate dust, despite their oxygen-rich nature. These results
corroborate our previous findings in $\omega$\,Centauri.

\end{abstract}


\vfill\eject 
\section{INTRODUCTION}
\label{sec:intro}

Stellar mass loss and dust production remain two of the most critical,
yet least understood, aspects of stellar evolution models, especially
in low-mass population II stars. High-mass Asymptotic Giant Branch
(AGB) stars (up to 8~$M_\odot$) evolve and lose mass on very short
timescales, resulting in short bursts of dust input into the
Interstellar Medium (ISM) following episodes of star formation.
Low-mass stars are more numerous than their high-mass counterparts and
live much longer, resulting in a more sustained dust input into the
ISM. Constraining these aspects is crucial for determining the impact
of low-mass stars -- the most numerous stars in the Universe -- on
galaxy evolution. Observations with the {\it Spitzer Space Telescope}
\citep{werner04,gehrz07} and the {\it AKARI} telescope \citep{murakami07}
have produced many studies of infrared photometry of resolved stellar
populations in globular clusters (GCs). Despite these new studies, the
sample of observations is still quite small, consisting of only a
handful of GCs and creating almost as many new questions as it has
answered.

\citet{origlia07} argued on the basis of their analysis of {\it
Spitzer} data for 47\,Tuc that dust-accompanied mass loss occurs along
the entire Red Giant Branch (RGB), down to the Horizontal Branch
(HB). However, this was not confirmed with {\it AKARI} observations
\citep{ita07}. In other GCs, dusty mass loss is only seen near the
very tip of the RGB \citep{boyer08,mcdonald09}.

Intracluster dust clouds should form as a result of dusty mass loss
from many stars during the time between Galactic plane
crossings. Searches for this dust in a large sample of GCs using {\it
Spitzer} and {\it AKARI} have resulted in no successful/significant
detections \citep{barmby09,matsunaga08}, with the exception of M15
\citep{boyer06}.  This lack of intracluster medium (ICM) dust in most
clusters provides a mystery as to the fate of dust produced by evolved
stars in GCs.

\subsection{NGC~362}
\label{sec:ngc362}

We present an infrared study of NGC~362, a bright GC in the southern
sky.  As one half of the classical ``second-parameter" GC pair with
NGC~288, NGC~362 is a relatively well-studied cluster
\citep[e.g.,][]{bellazzini01, catelan01}.  The wide availability of
multi-wavelength and astrometric data, along with the cluster's large
coeval stellar population, a metallicity intermediate to 47\,Tuc and
the bulk population of $\omega$\,Cen \citep[$\rm{[Fe/H]}$ =
$-$1.16 for NGC~362,][]{harris96}\footnotemark\footnotetext{The \citet{harris96} catalog was update in Februrary 2003. See http://www.physwww.physics.mcmaster.ca/$\sim$harris/mwgc.dat}, and proximity to us
\citep[8.5~kpc,][]{harris96} make it an ideal candidate for a study of
dusty mass loss on the AGB and upper RGB.

Compared to NGC~288, NGC~362 has a red HB, despite having a very
similar metallicity to the former (i.e., the ``first parameter'').
There are several theories for this discrepancy, including (and
possibly combining) cluster age, helium abundances, and RGB mass loss,
which may be affected by the central concentration of the cluster or
other environmental factors \citep{catelan01}. A recent study also
suggests environmental conditions during formation as a
second-parameter candidate \citep{fraix09}.

NGC~362 appears to be devoid of an ICM.  \citet{barmby09} find an
upper limit of $6.3 \times 10^{-5}~M_\odot$ of dust, more than two
orders of magnitude less than predicted based on the number of HB
stars and the time since the last Galactic plane-crossing ($\tau_{\rm
c}$, see Table~\ref{tab:props}). In addition, \citet{grindlay77} and
\citet{hesser77} searched for and found no ionized intracluster
medium. Despite the apparent lack of material collected from
mass-losing stars in the cluster, two studies have identified a small
population of sources with infrared excesses attributed to circumstellar
dust and mass loss \citep{origlia02,ita07}, and \citet{mcdonald07}
have identified cluster stars with H$\alpha$ profiles that indicate
mass loss in the absence of dust.

%%%%%%%%%%%%%%%%%%%%%%%%%%%%%%%%%%%%%%%%%%%%%%%%%%%%%%%%%%%%%%%
% Table 1:
\begin{deluxetable}{lcc}
\tablewidth{0pc}
\tabletypesize{\normalsize}
\tablecolumns{3} 
\tablecaption{NGC~362 Properties\label{tab:props}}

\tablehead{\colhead{Parameter}&\colhead{Value}&\colhead{Source}}
\startdata

Right Ascension (J2000)\dotfill     & 01$^{\rm h}$03$^{\rm m}$14\fs27 &   \\
Declination (J2000)\dotfill         & $-$70\degr 50\arcmin 53\farcs6  &   \\
Distance (kpc)\dotfill                     & 8.5                  & 3 \\
$\rm{[Fe/H]}$\dotfill                      & $-$1.16              & 3 \\
$M$ ($M_\odot$)\dotfill                    & 3.78 $\times$ 10$^5$ & 2 \\
$R_{\rm core}$ (\arcmin)\dotfill           & 0.17                 & 3 \\
$R_{\rm half\,mass}$ (\arcmin)\dotfill     & 0.81                 & 3 \\
$E(B-V)$ (mag)\dotfill                     & 0.05                 & 3 \\
($m-M$)$_{\rm V}$ (mag)\dotfill            & 14.65                & 3 \\
$\tau_{\rm c}$ (yr)\tablenotemark{a}\dotfill & 3 $\times$ 10$^7$    & 4 \\
$N_{\rm HB}$\dotfill                       & $1.6 \times 10^2$    & 1 \\
$v_{\rm esc,0}$ (km\,s$^{-1}$)\tablenotemark{b}\dotfill  & 46.7                 & 5 \\
$M_{\rm V}$ (mag)\dotfill                  & $-$8.35              & 3 \\
$L_{\rm bol}$ ($L_\odot$)                  & 4.54 $\times$ 10$^5$ & 3 \\
Heliocentric radial velocity (km\,s$^{-1}$)& 223.5 $\pm$ 0.5      & 3 
\enddata

\tablecomments{ \ Sources: (1) \citet{barmby09}, (2) \citet{gnedin97},
(3) \citet{harris96}, (4) \citet{odenkirchen97}, (5)
\citet{mclaughlin05}.}

\tablenotetext{a}{ \ $\tau_{\rm c}$ is the time since the last Galactic plane-crossing.}

\tablenotetext{b}{ \ $v_{\rm esc,0}$ is the escape velocity at $R = 0$.}

\end{deluxetable}
%%%%%%%%%%%%%%%%%%%%%%%%%%%%%%%%%%%%%%%%%%%%%%%%%%%%%%%%%%%%%%%

\section{OBSERVATIONS AND DATA REDUCTION}
\label{sec:obs}
 
\begin{figure} %ngc362_3color_1b4g5r.eps
\epsscale{1.2} \plotone{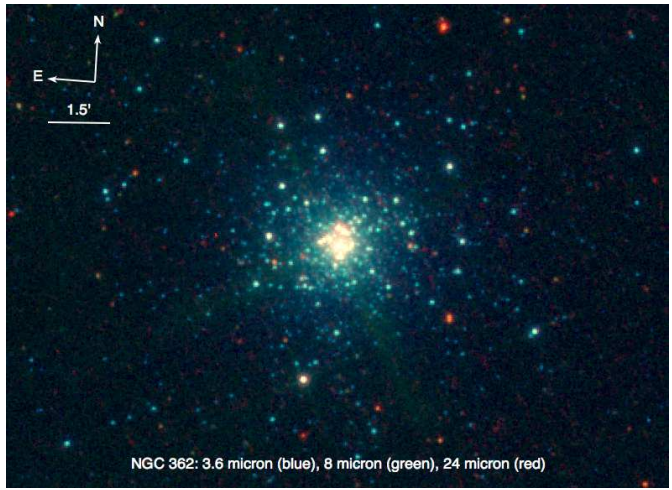} \figcaption{3-color IRAC
and MIPS image of NGC 362. Blue is 3.6~\micron{}, green is
8~\micron{}, and red is 24~\micron{}. The northern most portion of the
SMC bar lies approximately 1\degr{} to the south. \label{fig:3color} }
\end{figure}

A 3-color image composed of 3.6, 8, and 24~\micron{} Infrared Array
Camera (IRAC) and Multiband Imaging Photometer (MIPS) images of
NGC~362 is presented in Figure~\ref{fig:3color}. The cluster's
near-juxtaposition with the Small Magellanic Cloud (SMC) resulted in
serendipitous observations of NGC~362 with {\it Spitzer} as part of
the Surveying the Agents of Galaxy Evolution {\it Spitzer} Legacy
Program (SAGE-SMC; K.D.Gordon 2009, in preparation).  NGC~362 lies
approximately 2\degr{} north from the center of the SMC bar, placing
it near the edge of our 5\degr{} $\times$ 5\degr{} {\it Spitzer}
observations.  The cluster was covered at 3.6, 4.5, 5.8, 8, 24, and 70
\micron{} to well outside of its half-mass radius of 0.81\arcmin{}
\citep{harris96}.  SAGE-SMC 160~\micron{} images did not include the
cluster. All data and analysis presented here are confined to within
6\arcmin{} of the cluster center to minimize contamination from SMC
sources.

Observations consist of two epochs of IRAC images separated by 3
months and one epoch of MIPS images.  The 1\,$\sigma$ sensitivities
are 2.3, 3.4, 19.1, 20.5, and 28.1 $\mu$Jy for 3.6, 4.5, 5.8, 8, and
24~\micron{}, respectively. Short exposure times (12~s)
ensure that saturation of cluster stars is not an issue. Angular
resolutions for IRAC wavelengths range from 1.7\arcsec{} at
3.6~\micron{} to 1.9\arcsec{} at 8~\micron{} and increase to
5.8\arcsec{} for MIPS 24~\micron{}. For more details regarding data
acquisition and reduction for the SAGE-SMC program, see K.D.Gordon, et
al. (2009, in preparation).

Photometry should be reasonably complete to well beyond the HB
($M_{\rm HB} \approx -1$~mag at 3.6~\micron{}), as demonstrated by a
steady increase in source counts to $M_{3.6\mu \rm m} \approx 2$~mag
in Figure~\ref{fig:lumfunc}. While photometric completeness tests have
not yet been performed on the SAGE-SMC data, the completeness limits
beyond the crowded inner 1\arcmin{} of the cluster are likely similar
to those for the SAGE Large Magellanic Cloud (SAGE-LMC) IRAC data
\citep{meixner06} since the observations were similarly designed. At
3.6~\micron{}, the average photometric completeness limit for SAGE-LMC
is $\approx$16~mag ($M_{3.6\mu \rm m} \approx 1.4$~mag for
NGC~362). The luminosity function in the inner 1\arcmin{} of the
cluster is truncated just below the HB (Fig.~\ref{fig:lumfunc}); this
is a result of incompleteness due to crowding.  The fact that the
luminosity function at brighter levels does not differ in the cluster
core from the field means that crowding is not important for stars
brighter than the HB.

Only three point sources are detected at 70~\micron{}, and only one of
these (designated here as s11) is also detected in IRAC.  This source
is discussed further in Section~\ref{sec:spitagbs}.
				   
%%%%%%%%%%%%%%%%%%%%%%%%%%%%%%%%%%%%%%%%%%%%%%%%%%%%%%%%%%%%%%%
% Fig 1: 
%%%%%%%%%%%%%%%%%%%%%%%%%%%%%%%%%%%%%%%%%%%%%%%%%%%%%%%%%%%%%%%

\subsection{Ancillary Data}
\label{sec:ancdat}

To aid in estimating the temperatures and luminosities of the cluster
stars, we collected literature photometry spanning the optical and the
near-IR. NGC~362 was observed in the Magellanic Clouds Photometric
Survey \citep[MCPS;][]{zaritsky02}, the Deep Near-Infrared Southern
Sky Survey \citep[DENIS;][]{epchtein97}, the Naval Observatory Merged
Astrometric Dataset \citep[NOMAD;][]{zacharias05} and the Two Micron
All Sky Survey \citep[2MASS;][]{skrutskie06}.

The MCPS catalog includes UBVI photometry, but does not cover the
inner $\approx$0.3\arcmin{} of the cluster core, where four candidate
mass-losing stars identified with {\it Spitzer} data reside (see
Section~\ref{sec:stars}). We obtained optical photometry of stars in
the cluster core from the DENIS and Nomad catalogs, which include
$I$-, $J$-, and $K_{\rm s}$-band photometry and $B$-, $V$-, and
$R$-band photometry, respectively.  2MASS includes JHK$_{\rm s}$ photometry and
probes to the cluster center. This collection of optical and near-IR
data complements the 3.6~\micron{} {\it Spitzer} data in both
completeness and angular resolution, with 93\% of the 3.6~\micron{}
sources matched in at least one of these four ancillary catalogs. The
remaining 7\% of 3.6~\micron{} sources without optical or near-IR
counterparts have a mean 3.6~\micron{} magnitude of 16.5~mag, with a
standard deviation of 1.3~mag, or three magnitudes fainter than the
HB. All of the sources investigated in detail in this study
(Section~\ref{sec:stars}) are detected in at least one near-IR or
optical catalog.

In addition, \citet{ita07} obtained infrared photometry at 2.4, 3.2,
4.1, 7.0, 9.0, 11.0, 15.0, 18.0, and 24~\micron{} with the Infrared
Camera \citep[IRC;][]{onaka07} onboard the {\it AKARI} telescope \citep[see
  Fig. 3 from][photometry obtained through private
  communication]{ita07}. IRC images have angular resolutions ranging
from 0.9\arcsec{} at 2.4~\micron{} to 9\arcsec{} at 24~\micron{},
making stellar blending a problem in the most crowded regions of the
cluster.  

NGC~362 was also observed at 24 and 70~\micron{} by
\citet{barmby09}. While none of the candidate mass-losing stars
identified in Section~\ref{sec:stars} are detected at 70~\micron{}, the
supplementary 24~\micron{} data are useful for estimating dust
compositions and mass-loss rates of individual stars (see
Section~\ref{sec:mlrs}).

A Hubble ACS image was obtained with permission from the ACS Survey of
Galactic globular clusters team \citep{sarajedini07}. The high
resolution of this image was helpful in determining when stellar
blending in {\it Spitzer} and {\it AKARI} images is potentially a problem.

\subsection{Cluster Membership}
\label{sec:mem}

Proper motions in NGC~362 were obtained by \citet{zacharias05} and
\citet{tucholke92}. The latter computed a membership likelihood by
fitting the distribution of stars in $\mu_\alpha$cos$\delta$,
$\mu_\delta$ space with a sum of two bivariate Gaussians representing
NGC~362 and field stars. We use these probability estimates to
eliminate non-members.  For proper motions from \citet{zacharias05},
we employ a conservative cut-off of $>$35~mas~yr$^{-1}$ to eliminate
probable cluster non-members.

The heliocentric radial velocity ($v_{\rm rad}$) of NGC~362 is
223.5~$\pm$~0.5~km~s$^{-1}$ \citep{harris96}.  \citet{fischer93}
measured radial velocity for 215 stars around the center of the
cluster, and confirmed two stars in their sample as radial velocity
non-members.  \citet{harris06} measure the radial velocities of red
giants in the SMC as $146~\pm~28~{\rm km~s}^{-1}$, and find that it is
rare for an SMC star to have a radial velocity larger than $200~{\rm
  km~s}^{-1}$. However, they did not measure stars at positions near
NGC~362, and we therefore note that it is possible, although rare, for
SMC stars to have radial velocities similar to NGC~362 due to the
rotation of the SMC (for relatively high-mass stars) and its velocity
dispersion (in particular for low-mass stars).

Membership information is available for 68\% of the sources detected by
{\it Spitzer}. We suspect that many of the sources without membership
information (especially those with extreme IR excess) are in fact
background galaxies that are not detected in the optical (see
Section~\ref{sec:spitagbs}).
%%%%%%%%%%%%%%%%%%%%%%%%%%%%%%%%%%%%%%%%%%%%%%%%%%%%%%%%%%%%%%%
% Fig 2: 
\begin{figure} %lumfunc_absmag.ps
\epsscale{1.2} \plotone{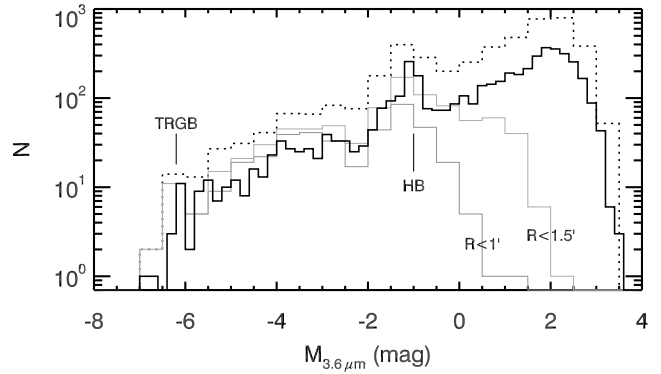} \figcaption{Luminosity function of
 NGC~362 (within $R=6$\arcmin), including both epochs of IRAC
 data. The solid and dotted lines illustrate 0.2 and 0.5 magnitude
 bins, respectively. The tip of the Red Giant Branch (TRGB) is located
 near $M_{3.6\mu \rm m}~\approx~-6.2$~mag, and the Horizontal Branch
 (HB) is located near $M_{3.6\mu \rm m}~\approx~-1.0$~mag. Photometry
 should be reasonably complete to approximately 3 magnitudes fainter
 than the HB at 3.6~\micron. In the inner 1\arcmin{} of the cluster
 (gray line) crowding limits source completeness to magnitudes
 brighter than the HB.\label{fig:lumfunc} }
\end{figure}
%%%%%%%%%%%%%%%%%%%%%%%%%%%%%%%%%%%%%%%%%%%%%%%%%%%%%%%%%%%%%%%

\section{Mass-Losing Stars in NGC~362}
\label{sec:stars}

\subsection{Previously Identified Mass-Losing Stars}
\label{sec:prevagbs}

Three sources in NGC~362 \citep[designated as x01, x02, and x03
by][]{mcdonald07} were identified as having infrared excess by
\citet{origlia02} using 12-\micron{} {\it Infrared Space Observatory} ({\it ISO}) data. The coordinates of
source x01 coincide with a {\it Spitzer} source that shows
24~\micron{} excess (source s04, see
Table~\ref{tab:agbs}). \citet{mcdonald07} showed x01 to have a red
H$\alpha$ line emission wing and blue-shifted line absorption core,
indicative of an outflow with a mass-loss rate of
$\sim$$10^{-6}~M_\odot~{\rm yr^{-1}}$.  Two other IR excessive point
sources (sources s03 and s05) lie immediately to either side of this
source, and while resolved in {\it Spitzer}, all three sources are
likely blended together in {\it ISO} images.

{\it ISO} source x02 corresponds to three different objects, two of which
(x02a and x02b) are detected with {\it Spitzer} and show little to no
IR excess despite having blue-shifted H$\alpha$ absorption cores
\citep{mcdonald07}. The third source (x02c) is unresolved by {\it
Spitzer}. Although apparently exhibiting a lack of IR excess, we note
that x02a and x02b are located in the most dense region of the cluster
where source confusion in {\it Spitzer} is severe and reliable
photometry is difficult.

{\it ISO} source x03 is not resolved at 24~\micron{} with MIPS, but its IRAC
colors do indicate a possible (slight) 8~\micron{} excess.  The
{\it AKARI} observation of x03 measures 3.2, 4.1, 7, and 9~\micron{} fluxes
that are four times brighter than the {\it Spitzer} 3.6, 4.5, and
8~\micron{} fluxes, likely due to stellar blending in {\it AKARI} since x03 is
located in a dense region of the cluster. Despite this discrepancy,
{\it AKARI} fluxes still show x03 to have a very slight IR excess. A strong
red emission wing in the star's H$\alpha$ line and blue-shifted
absorption core yield a potential mass-loss rate of
$\sim$$10^{-6}~M_\odot~{\rm yr^{-1}}$ \citep{mcdonald07}. The lack of
reliable photometry redward of 8~\micron{} prevents us from measuring
a mass-loss rate in Section~\ref{sec:mlrs}.

In addition to the three {\it ISO} sources, \citet{mcdonald07} obtained
VLT/UVES spectra of 10 other stars in NGC~362, all of which have
spectra typical for oxygen-rich red giant stars.  Five of these
sources show IR excess at 24~\micron{} (see Table~\ref{tab:agbs} and
Section~\ref{sec:mlrs}). Fitting of the H$\alpha$ profiles suggests
mass-loss rates ranging from $10^{-7} - 10^{-5}~M_\odot~{\rm
yr^{-1}}$. Source o01 ({\it Spitzer} source s06) has strong IR excess,
is variable \citep[V16, $P = 138$ days;][]{clement97}, and shows
strong molecular bands and emission in H$\alpha$.

The remaining five sources from \citet{mcdonald07} are also detected
in IRAC. None of these sources show significant 8 or 24~\micron{}
excess, the exception being o09, which shows slight 8~\micron{} excess
($[3.6]-[8] = 0.18$, $M_{\rm 3.6 \mu m} = -5.1$~mag). It is
interesting to note that all sources observed by \citet{mcdonald07}
that show IR excess in {\it Spitzer} or {\it ISO} data also show red
H$\alpha$ line emission wings. The four stars (o02, o03, o04, and o10)
that have no IR excess show only blue H$\alpha$ line emission
wings. \citet{mcdonald07} suggest that stars showing only blue
emission wings have heated chromospheric material restricted to radii
less than $2~R_\odot$, resulting in the red emission being blocked by
the stellar disk. In this scenario, the chromospheres of stars showing
significant IR excess will be extended to large enough radii to see a
red emission in H$\alpha$.

\citet{smith99} obtained an optical spectrum of another NGC~362
variable from \citet{clement97}, V2 ($P = 90$ days). V2 is {\it
Spitzer} source s02 (Table~\ref{tab:agbs}, Section~\ref{sec:extreme}),
which has strong IR excess. Strong H$\alpha$ and H$\beta$ emission in
V2 indicate chromospheric activity, and a strong
\ion{Li}{1}\,$\lambda$6707 feature indicates a lithium
overabundance. Low-mass giant stars have typically destroyed and
diluted their photospheric lithium (which is brought to the surface
via the first dredge-up), so those showing a lithium overabundance are
rare, although not unheard of.

Eight sources in NGC~362 showing strong IR excess (F$_{24\mu \rm
m}/$F$_{7\mu \rm m}~>~1$, where zero-excess stars have F$_{24\mu \rm
m}/$F$_{7\mu \rm m}~\approx~0.1$ ) were identified by \citet{ita07}
using {\it AKARI} data.  These eight sources are located in the region of
the IR color-magnitude diagram (CMD) that is now attributed to
background galaxies
\citep[e.g.,][]{blum06,boyer08,boyer09,bolatto07,mcdonald09}, although
there may also be a small amount of contamination from background
evolved stars in this region. Several of the cluster members showing
IR excess in {\it Spitzer} data are also detected by \citet{ita07},
but these sources are not the focus of that particular study.

\subsection{Identifying Mass-Losing Stars with Spitzer}
\label{sec:spitagbs}

\begin{figure} %ngc362_cmd45.ps
\epsscale{1.2} \plotone{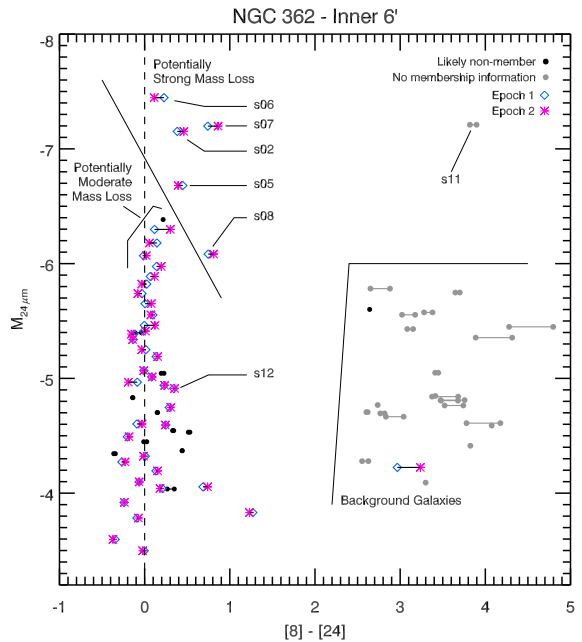} \figcaption{$[24]$ versus
$[8] - [24]$ color-magnitude diagram.  Lines connect points from each
IRAC epoch. Five sources are identified as candidates of strong dusty
mass-loss, and five others are candidates for moderate dusty mass
loss. One of the moderately dusty candidates is a likely cluster
non-member (source s01). \label{fig:cmd45} }
\end{figure}

A CMD showing {\it Spitzer} $[24]$ versus $[8]-[24]$ is presented in
Figure~\ref{fig:cmd45}. Confirmed non-members are represented with
black dots, and sources without membership information are marked with
gray dots, the majority of which are likely background galaxies. Stars
on the RGB have average colors of $[8]-[24]~=~0$~mag, while stars near
the bright end of the RGB begin to shift to $[8]-[24]~>~0$~mag.

Table~\ref{tab:agbs} lists ten sources with $[8]-[24]~>~0$~mag located
in the region of the CMD above or near the tip of the Red Giant Branch
(TRGB; $M_{3.6\mu \rm m}~\approx~-6.2$~mag, based on the drop-off in
the luminosity function, shown in Figure~\ref{fig:lumfunc}). Proper
motion from \citet{tucholke92} indicates that source s01 is a likely
non-member, but the remaining nine sources are confirmed radial
velocity or proper motion cluster members. Five of the ten sources
(s02, s05, s06, s07, and s08) are either very bright or have strong
24~\micron{} excess, and we designate them as candidate strong
mass-losing stars. The four more moderate confirmed member stars are
potentially forming dust in smaller quantities and possibly also
losing mass at a moderate rate.

%%%%%%%%%%%%%%%%%%%%%%%%%%%%%%%%%%%%%%%%%%%%%%%%%%%%%%%%%%%%%%%
% Table 2:
\begin{deluxetable*}{cccccccl}
\tablewidth{6in}
\tabletypesize{\footnotesize}
\tablecolumns{8} 
\tablecaption{Candidate Strong and Moderate Mass-losing Stars in NGC~362\label{tab:agbs}}

\tablehead{\colhead{Source}&\colhead{R.A. (J2000)}&\colhead{Dec. (J2000)}&\colhead{M$_{8\mu
\rm m}$}&\colhead{M$_{24\mu \rm
m}$}&\colhead{T}&\colhead{L}&\colhead{Notes and}\\\colhead{ID}&\colhead{(h m s)}&\colhead{(\degr~\arcmin~\arcsec)}&\colhead{(mag)}&\colhead{(mag)}&\colhead{(K)}&\colhead{($L_\odot$)}&\colhead{Alternate Designations\tablenotemark{a}}}
\startdata

\vspace{-0.2em}\\

\multicolumn{8}{c}{Moderate Mass-Loss Candidates}\\

\hline

% 6926
s01&01 03 35.75&$-$70 50 52.4&$-$6.17$\pm$0.03&$-$6.38$\pm$0.02&3950&1656&non-member\\
% 9612
s03&01 03 20.07&$-$70 50 55.0&$-$6.09$\pm$0.03&$-$6.30$\pm$0.02&4339&2184&o07, post-AGB?\\
% 9814
s04&01 03 19.07&$-$70 50 51.1&$-$6.07$\pm$0.03&$-$6.07$\pm$0.03&3823&1551&x01\\
%11278
s09&01 03 10.67&$-$70 50 54.4&$-$6.08$\pm$0.03&$-$6.18$\pm$0.03&4226&1932&o08, post-AGB?\\
%16482
s10&01 02 43.34&$-$70 48 47.8&$-$5.81$\pm$0.03&$-$5.98$\pm$0.03&3975&1363&\\

\hline

\vspace{-0.2em}\\

\multicolumn{8}{c}{Strong Mass-Loss Candidates}\\

\hline

% 8410
s02&01 03 21.85&$-$70 54 20.2&$-$6.73$\pm$0.03&$-$7.15$\pm$0.02&3907&1826&V2 ($P=90$ days)\\
%10125
s05&01 03 17.29&$-$70 50 49.1&$-$6.26$\pm$0.05&$-$6.68$\pm$0.02&4058&2280&o05a/o05b\\
%10574
s06&01 03 15.08&$-$70 50 31.5&$-$7.28$\pm$0.03&$-$7.45$\pm$0.02&3962&3106&o01, V16 ($P=138$ days)\\
%10830
s07&01 03 13.62&$-$70 50 36.5&$-$6.40$\pm$0.03&$-$7.20$\pm$0.02&3343&1402&o06\\
%10885
s08&01 03 12.60&$-$70 51 00.9&$-$5.31$\pm$0.02&$-$6.08$\pm$0.02&3682&1842&stellar blend?

\enddata

\tablecomments{ \ The prefix of the source IDs is short for ``{\it Spitzer}''. Sources are numbered in order of decreasing R.A.}

\tablenotetext{a}{ \ Designations from \citet{mcdonald07} have a
prefix ``o'' or ``x'', followed by a number. Source s04 (x01) was also
detected by \citet{origlia02}. Photosphere temperatures derived here
for s03 and s09 are warm, possibly indicating that these are post-AGB
stars. However, \citet{mcdonald07} estimate temperatures that place
both stars firmly on the RGB.}

\end{deluxetable*}
%%%%%%%%%%%%%%%%%%%%%%%%%%%%%%%%%%%%%%%%%%%%%%%%%%%%%%%%%%%%%%%

%%%%%%%%%%%%%%%%%%%%%%%%%%%%%%%%%%%%%%%%%%%%%%%%%%%%%%%%%%%%%%%
% Fig 3: 
%%%%%%%%%%%%%%%%%%%%%%%%%%%%%%%%%%%%%%%%%%%%%%%%%%%%%%%%%%%%%%%

\begin{figure} %ngc362_cmd14.ps
\epsscale{1.2} \plotone{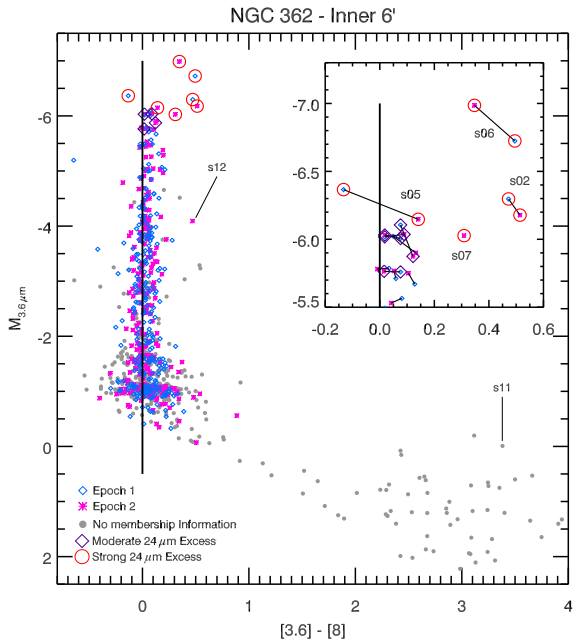} \figcaption{$[3.6]$ versus
$[3.6]-[8]$ color-magnitude diagram. Lines connect points from each
IRAC epoch. The sources identified as candidate stars with strong and
moderate mass loss in Figure~\ref{fig:cmd45} are marked with red
circles and purple diamonds, respectively. The crowded region towards
the TRGB is inset at the top right. The source at $[8] - [24] > 3.5$
and $M_{24\mu \rm m} < -7$~mag in Figure~\ref{fig:cmd45} (source s11)
is located far from mass-losing AGB stars and near unresolved
background galaxies in this CMD. Source s12 may be a background SMC
carbon star. \label{fig:cmd14} }
\end{figure}

\begin{figure} %sed_16466.ps
\epsscale{1.2} \plotone{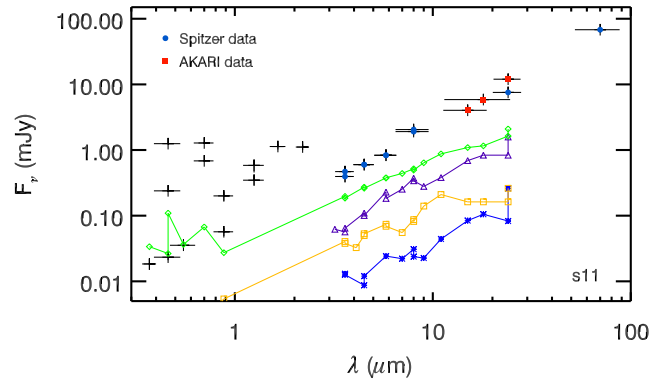} \figcaption{SED of source s11, a
suspected background galaxy. Black pluses, blue circles, and red
circles represent optical/Near-IR, {\it Spitzer}, and {\it AKARI} data,
respectively. Wavelength error bars correspond to the filter widths.
The scatter in the optical and near-IR photometry is likely due to
mis-matching of sources, as foreground optical sources may lie
directly along the sightline to the galaxy. Four other sources from
the galaxy locus in Figure~\ref{fig:cmd45} are also shown (green
diamonds, purple triangles, orange squares, and blue asterisks),
plotted with offset fluxes for visual clarity. The SED of s11
resembles that of the other four background galaxies.
\label{fig:galsed} }
\end{figure}

In addition to the ten sources listed in Table~\ref{tab:agbs}, source
s11 (R.A. = 1$^{\rm h}$02$^{\rm m}$48\fs66, Dec. =
$-$70\degr45\arcmin22\farcs6) is very red and bright, with
$[8]-[24]~>~3.5$ and $M_{24\mu \rm m}~<~-7$~mag. This source is the
only IRAC source also detected at 70~\micron{}, and it is most likely
an unresolved background galaxy, despite being more than a magnitude
brighter than the general locus of other unresolved background
galaxies. Its position in the $[3.6]$ versus $[3.6]-[8]$ CMD
(Fig.~\ref{fig:cmd14}) and its inferred bolometric luminosity if it
were a cluster member (see Section~\ref{sec:hrd} and
Figure~\ref{fig:hrd}) place it among the other background galaxies,
far from the mass-losing AGB stars. Moreover, this source's SED
(Fig.~\ref{fig:galsed}) rises strongly beyond 3~\micron{}, resembling
those of galaxies in the SINGS sample \citep[e.g.,][]{dale06} and of
other background galaxies identified in Figure~\ref{fig:cmd45}.  While
we cannot confirm this source's status as a background galaxy without
a spectrum or high-resolution image (S11 falls outside of the ACS
field of view), we proceed under the assumption that it is not a
member of NGC~362 and exclude it from our analysis.

%%%%%%%%%%%%%%%%%%%%%%%%%%%%%%%%%%%%%%%%%%%%%%%%%%%%%%%%%%%%%%%
% Fig 4: 
%%%%%%%%%%%%%%%%%%%%%%%%%%%%%%%%%%%%%%%%%%%%%%%%%%%%%%%%%%%%%%%

%%%%%%%%%%%%%%%%%%%%%%%%%%%%%%%%%%%%%%%%%%%%%%%%%%%%%%%%%%%%%%%
% Fig 5: 
%%%%%%%%%%%%%%%%%%%%%%%%%%%%%%%%%%%%%%%%%%%%%%%%%%%%%%%%%%%%%%%

It is clear from the {\it Spitzer} data that dusty mass loss in
NGC~362 is confined to the upper RGB/AGB. There {\it are} a handful of
potential member stars with moderate 8~\micron{} excess at
$-2\,\gtrsim\,M_{3.6\mu \rm m}\,\gtrsim\,-4$~mag visible in
Figure~\ref{fig:cmd14}. All of these sources are very near the
cluster center, and upon visual inspection, all but one are clearly
blended with other sources in all of the {\it Spitzer} data,
potentially creating artificially red $[3.6]-[8]$ colors if the flux
is not accurately extracted at one or both wavelengths. Indeed it is
likely that the flux of these sources is inaccurate since these are
only moderately bright stars located in an extremely crowded region of
the cluster. Given that the mean uncertainty in their $[3.6]-[8]$
colors is 0.17~mag ($0.2 \gtrsim [3.6]-[8] \lesssim 0.3$~mag), these
sources could easily be members of the non-dusty RGB.

\begin{figure} %sed_11261.ps
\epsscale{1.2} \plotone{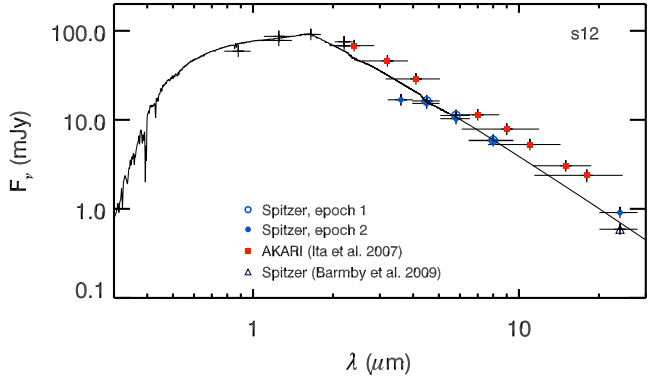} \figcaption{SED of source
s12. The solid black line shows the best fit {\sc marcs} model (see
Section~\ref{sec:hrd}). The underluminous 3.6~\micron{} flux appears
to be the cause of the $[3.6]-[8]$ excess apparent in
Figure~\ref{fig:cmd14}, and may be due to poor photometry or possibly
to absorption from acetylene. {\it AKARI} fluxes are systematically too
bright, which may be due to stellar blending. \label{fig:11261} }
\end{figure}

\begin{figure*} %myhrd.ps
\epsscale{1} \plotone{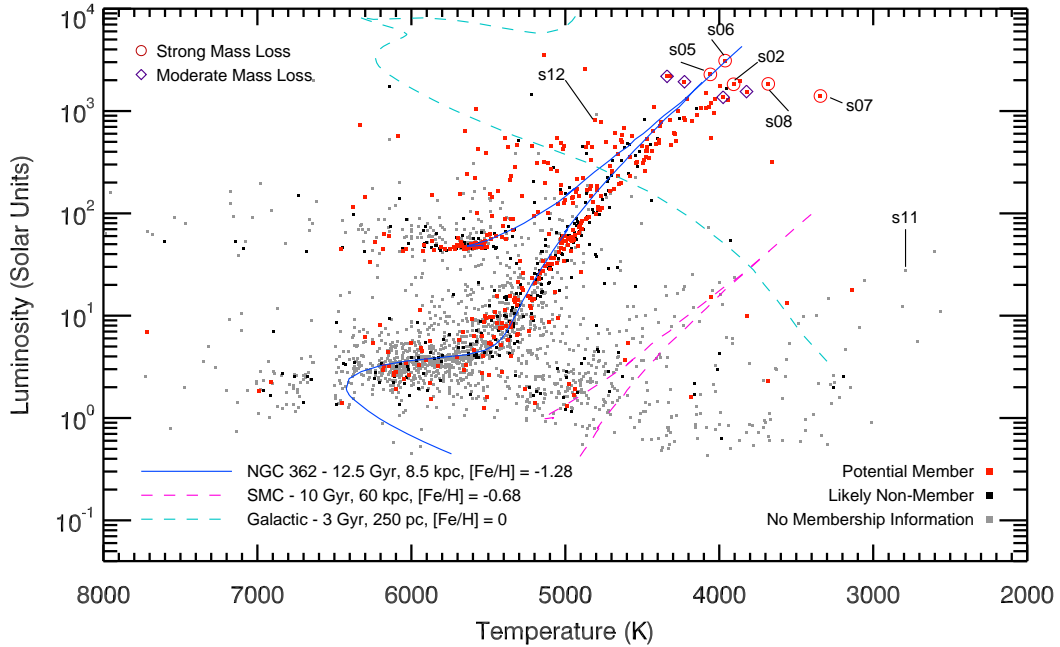} \figcaption{Physical HRD for
NGC~362. As in Figure~\ref{fig:cmd14}, candidate strong mass-losing
stars are circled in red, and candidate moderate mass-losing stars
(excluding s01, a likely cluster nonmember) are marked with purple
diamonds (see Table~\ref{tab:agbs}). Padova isochrones are also
plotted in dark blue for NGC~362, light blue (dashed line) for
Galactic stars at a representative distance, and magenta (dashed line)
for the SMC. Note that the RGB is cooler than expected.\label{fig:hrd}
}
\end{figure*}

The only potentially unblended source in this region of the CMD (s12;
R.A. = 1$^{\rm h}$03$^{\rm m}$11\fs59, Dec. =
$-$70\degr50\arcmin23\farcs5) is located at $[3.6]-[8]~\approx~0.47 \pm 0.14$
and $M_{3.6\mu \rm m}~\approx~-4$~mag in Figure~\ref{fig:cmd14}. This
source was designated a radial velocity member by \citet{fischer93}
($v_{\rm rad}~=~224.8~{\rm km~s}^{-1}$), but it has no known proper
motion measurements.  IR data for this source are somewhat ambiguous
(Fig.~\ref{fig:11261}), and it is very likely that either the IRAC
photometry is inaccurate or this source is actually a background star
belonging to the SMC. The source is detected at 3.6~\micron{} in only
epoch two of the IRAC data, and this flux falls well below the
apparent stellar continuum, resulting in what may be an artificially
red $[3.6]-[8]$ color.

%%%%%%%%%%%%%%%%%%%%%%%%%%%%%%%%%%%%%%%%%%%%%%%%%%%%%%%%%%%%%%%
% Fig 6: 
%%%%%%%%%%%%%%%%%%%%%%%%%%%%%%%%%%%%%%%%%%%%%%%%%%%%%%%%%%%%%%%

Despite the under-luminous 3.6~\micron{} flux and systematically
over-luminous {\it AKARI} data, the SAGE-SMC 24~\micron{} measurement of
source s12 does show a very slight amount of excess. It is possible
that this source may be a carbon star belonging to the SMC, and the
discrepant 3.6~\micron{} flux may indicate variability or possibly
molecular absorption from acetylene \citep[cf.][]{vanloon08}. The
heliocentric radial velocity for carbon stars on the outskirts of the
SMC is $149.3~\pm~30~{\rm km~s}^{-1}$, with a dispersion of $\delta
v~=~25.2~\pm~2.1~{\rm km~s}^{-1}$ \citep{hatzidimitriou97}. A total of
6\% of these carbon stars have velocities $200~{\rm
km~s}^{-1}~<~v_{\rm rad}~<~250~{\rm km~s}^{-1}$.  Given that this
source has a magnitude and color very similar to typical carbon stars
in the SMC \citep{bolatto07}, it is likely not a member of NGC~362,
and, as a consequence, no significant dusty mass loss occurs below the
TRGB in NGC~362.

%%%%%%%%%%%%%%%%%%%%%%%%%%%%%%%%%%%%%%%%%%%%%%%%%%%%%%%%%%%%%%%
% Fig 7: 
%%%%%%%%%%%%%%%%%%%%%%%%%%%%%%%%%%%%%%%%%%%%%%%%%%%%%%%%%%%%%%%

\subsection{The Hertzsprung-Russell Diagram}
\label{sec:hrd}

In Figure~\ref{fig:hrd}, we present a physical Hertzsprung-Russell
Diagram (HRD) for NGC~362.  To determine the stellar parameters of all
stars in the cluster, we fit optical, near-IR, and IRAC photometry to
the Model Atmosphere in Radiative and Convective Scheme ({\sc marcs}) code
\citep{gustafsson75, gustafsson08}. The grid of model spectra and
fitting technique used here are described in more detail in
\citet{mcdonald09}. To fit the model spectra to the data, we give
preference to MCPS photometry, including NOMAD and DENIS photometry
only if a source is not present in the MCPS catalog.

A Padova isochrone \citep{marigo08} is included in
Figure~\ref{fig:hrd} at 12.5~Gyr and $\rm{[Fe/H]}~=~-1.28$. This
metallicity is slightly lower than the $\rm{[Fe/H]}~=~-1.16$ quoted in
\citet{harris96}, which we use to calculate mass-loss rates in
Section~\ref{sec:mlrs}. We note that a value of $\rm{[Fe/H]} = -1.28$
would result in a 32\% increase in the gas-to-dust ratio ($\psi$) and
therefore a 15\% decrease in the dust mass loss. Also plotted are
isochrones representing the SMC and distant Galactic stars. As in
$\omega$\,Cen \citep{mcdonald09}, the NGC~362 upper RGB is cooler than
the Padova isochrone, which may suggest (dusty or non-dusty) mass loss
on the RGB. RGB mass loss is not well accounted for in Simple Stellar
Population models applied to distant galaxies, but it may be of
critical importance in predicting effective temperatures and thus
colors and bolometric corrections \citep{salaris05}.

The brightest/dustiest candidate mass-losing stars in the cluster are
circled in red in Figure~\ref{fig:hrd}, and the moderately dusty stars
are marked with purple diamonds. The very red source directly above
the galaxy locus in Figure~\ref{fig:cmd45} (source s11) is
underluminous and very cold on the HRD, consistent with the assertion
that it is a background galaxy. Note that two of the moderately dusty
sources are slightly warmer than the RGB (sources s03 and s09), which
could indicate that these are post-AGB stars, although
\citet{mcdonald07} indicated photosphere temperatures of $\approx
3900$~K for both stars, placing them firmly on the RGB. Source s07 is
the coolest of the mass-losing stars at $T \approx
3300$~K. \citet{mcdonald07} measure a temperature of $T \approx
3950$~K, placing s07 among the other strong mass-losing stars.  All
other candidate mass-losing stars have typical temperatures and
luminosities.

There is an anomalous source in Figure~\ref{fig:hrd} with $T \approx
3700~{\rm K}$ and $L \approx 300~L_\odot$. If this source is at the
distance of the SMC, its luminosity is $L \sim 15\,000~L_\odot$ ,
which is typical for a carbon star. Source s12, which we suggested
may also be an SMC carbon star in Section~\ref{sec:spitagbs}, was fit to a
{\sc marcs} model with $T \approx 4800~{\rm K}$ and $L \approx 820~L_\odot$
(see Fig.~\ref{fig:11261}). However, we note that a lack of photometry
shortward of 0.9~\micron{} for source s12 resulted in an uncertain
fit.

\subsection{Distribution of Dusty Stars}
\label{sec:dist}

\begin{figure} %dist.ps
\epsscale{1.15} \plotone{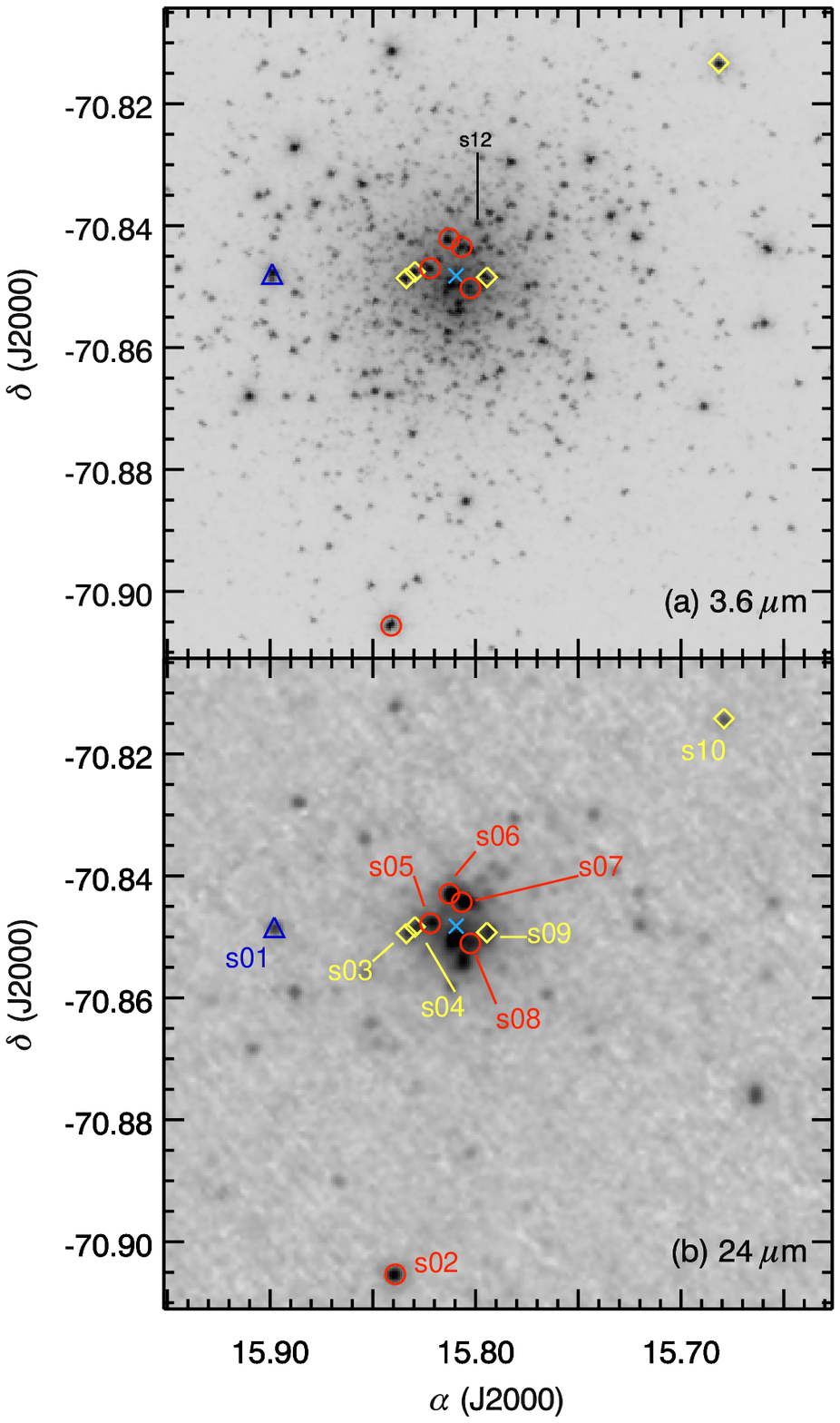} \figcaption{Distribution of
candidate mass-losing stars.  Most dusty stars are located within the
inner 1\arcmin{} of the cluster center. Red circles mark sources with
strong mass loss, and yellow diamonds mark sources experiencing
moderate mass loss.  The open blue triangle (source s01) is a likely
cluster non-member. Source s11, a background galaxy, falls well
outside the field of view shown here.\label{fig:dist} }
\end{figure}

Figure~\ref{fig:dist} shows the spatial positions of the ten sources
identified as candidate dusty mass-losing stars (including s01, a
likely cluster nonmember).  Within the inner 3.5\arcmin{} of the
cluster, which includes all of the candidate dusty stars, confirmed
cluster members are a mean distance of 1.2\arcmin{} $\pm$ 0.1\arcmin{}
to the cluster center. For the dusty stars, this is 0.7\arcmin{} $\pm$
0.2\arcmin{}. While the dusty stars are slightly more tightly
concentrated than the general population, the uncertainties easily
allow for the possibility that both populations have the same central
concentration, supporting the assertion that these stars are indeed
cluster members.

Low-mass stars lose most of their mass on the RGB, and mass
segregation will cause AGB stars to migrate to the outer regions of
the cluster given that the time between the RGB and AGB phases is on
the order of the relaxation time scale of NGC~362
\citep[$\sim$$10^8$~yr;][]{fischer93}.  However, most of the dusty
stars in NGC~362 are likely RGB stars at the TRGB, and have therefore
not had enough time for mass loss to alter their orbits, unless
significant dustless mass loss occurred much earlier. We therefore do
not expect to find that the dusty stars have orbits differing from
those of the bulk population.

%%%%%%%%%%%%%%%%%%%%%%%%%%%%%%%%%%%%%%%%%%%%%%%%%%%%%%%%%%%%%%%
% Fig 8: 
%%%%%%%%%%%%%%%%%%%%%%%%%%%%%%%%%%%%%%%%%%%%%%%%%%%%%%%%%%%%%%%

\subsection{Mass-Loss Rates}
\label{sec:mlrs}

\begin{figure*}  %seds_mod.ps
\epsscale{1} \plotone{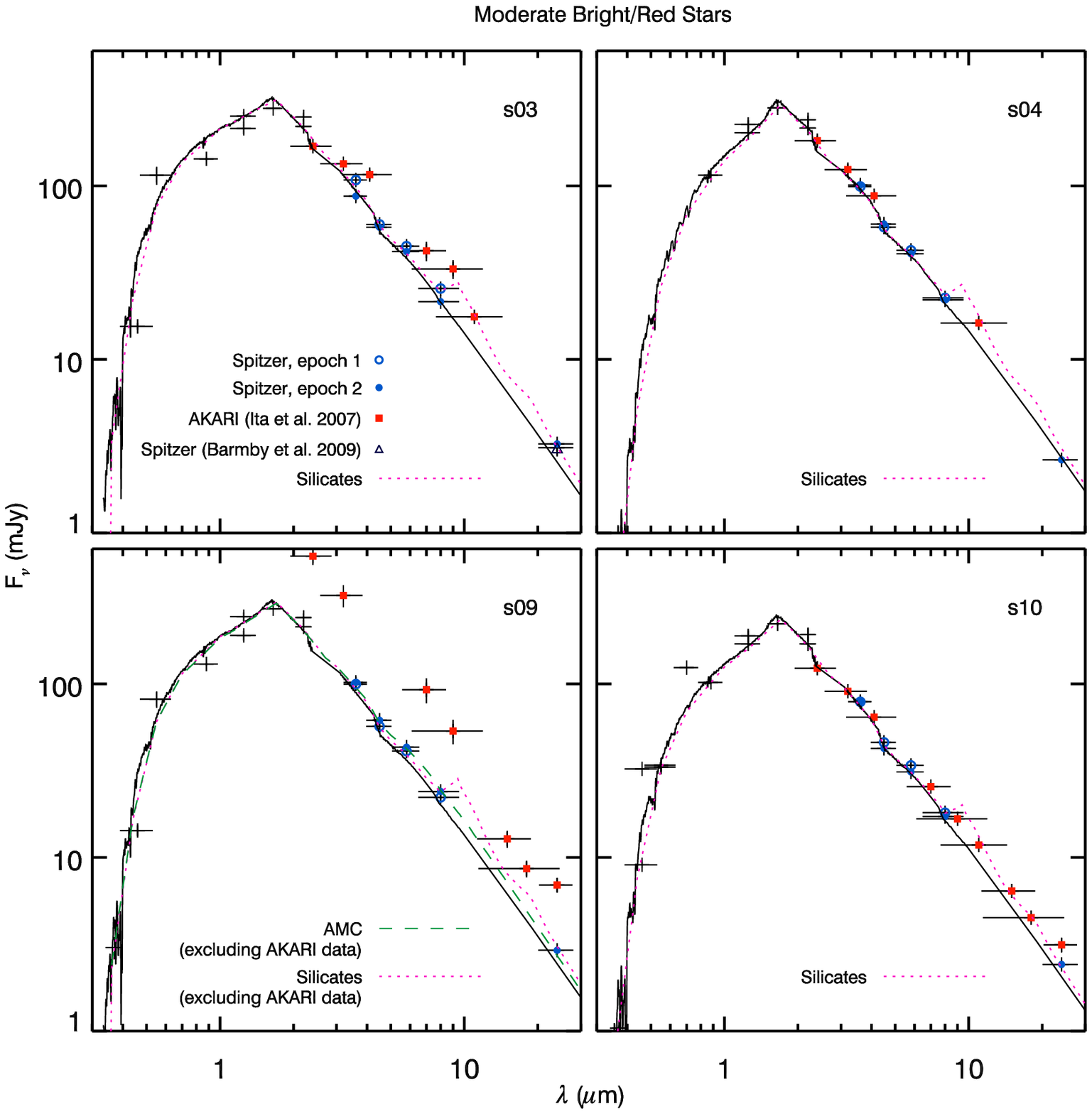} \figcaption{SEDs of moderately
dusty stars at the TRGB in NGC~362. Wavelength error bars correspond
to the filter width.  {\it AKARI} data are plotted as red squares and
{\it Spitzer} data are plotted as blue circles. {\sc marcs} models are
solid black lines. IRAC epoch~1 is plotted with open blue circles, and
IRAC epoch~2 is plotted with closed blue circles. All four stars are
well fit by a {\sc dusty} silicate model (magenta, dotted line), but
s09 is fit equally well by a {\sc dusty} amorphous carbon (AMC) model
(green, dashed line).\label{fig:seds_mod} }
\end{figure*}

\begin{figure*} %seds_strong.ps
\epsscale{1} \plotone{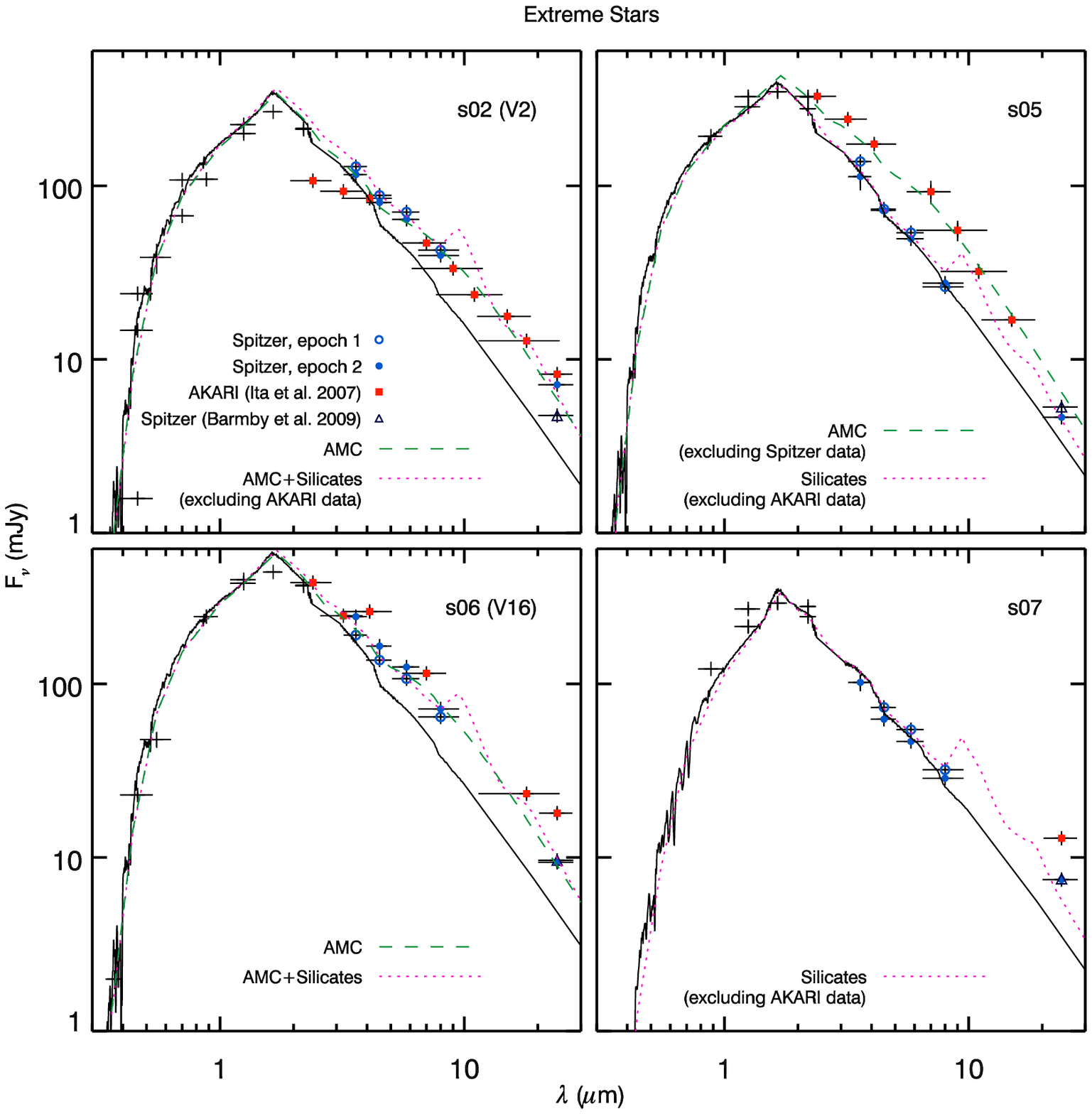} \figcaption{Same as
Figure~\ref{fig:seds_mod}, but for the most extreme stars in
NGC~362. The amorphous carbon (AMC) $+$ silicate {\sc dusty} models
shown for s02 and s06 are a combination of 50\% AMC and 50\%
silicates.\label{fig:seds_strong} }
\end{figure*}

The SEDs of the moderately dusty stars are presented in
Figure~\ref{fig:seds_mod}, and the SEDs of the more extreme stars are
presented in Figure~\ref{fig:seds_strong}. {\it AKARI} data
\citep{ita07} are plotted as red squares, {\it Spitzer} data are
plotted as blue circles, pluses indicate literature optical
photometry, and best-fit {\sc marcs} spectra are plotted with solid
lines. None of the moderate stars appears strongly variable between
{\it Spitzer} epochs and the {\it AKARI} epoch ({\it AKARI} data for
source s09 is unreliable). The extreme stars do show potential
moderate variability.

The {\it AKARI} data for sources s09 and s07 are very inconsistent with the
{\it Spitzer} data, which may be the result of source confusion in
{\it AKARI}. Without exception, {\it AKARI} 24~\micron{} fluxes are bright
compared to {\it Spitzer} 24~\micron{} photometry. This discrepancy is
probably due to the poor angular resolution of {\it AKARI} compared to {\it
Spitzer} (see Section~\ref{sec:ancdat}), resulting in source confusion
and blending. All but three of the sources in Table~\ref{tab:agbs}
were also detected at 24~\micron{} by \citet{barmby09}.  These
24~\micron{} data are very consistent with the {\it Spitzer} SAGE-SMC
24~\micron{} data, typically differing by a tenth of a magnitude or
less. Source s02 is the only exception to this, with a difference of
0.5~mag between {\it Spitzer} SAGE-SMC and \citet{barmby09}.

To estimate the mass-loss rates, we fit the SEDs using the {\sc dusty}
modeling code \citep{nenkova99}. Following \citet{mcdonald09}, the
SEDs are fit by models using dust composed of either amorphous carbon
\citep[AMC,][]{hanner88} or silicates, with the latter composed of
65\% astronomical silicates \citep{draine84}, 15\% compact Al$_2$O$_3$
(optical constants from the Jena database\footnotemark
\footnotetext{http://www.astro.uni-jena.de/Users/database/entry.html}),
and 10\% each of glassy and crystalline silicates \citep{jager94}.
These two dust compositions result in good fits, but the dust may
consist of other silicate or carbon species or different proportions
of species. Carbonaceous dust yields relatively blue $[8]-[24]$
colors, whereas oxygen-rich dust always produces a relatively large
24~\micron{} excess compared to 8~\micron{}, although without {\it AKARI}
data to fill the gap between 8 and 24~\micron{} left by {\it Spitzer},
it is difficult to definitively distinguish between different dust
compositions.

Following \citet{mcdonald09}, the resulting mass-loss rates from {\sc
dusty} were converted to real mass-loss rates using the following
prescription, which implicitly assumes a scaling of the wind speed
based upon a dust-driven wind formalism:

\begin{equation}
\dot{M}_{\rm dust} = \frac{\dot{M}_{\rm DUSTY}}{200} \left(\frac{L}{10^4}\right)^{3/4} \left(\frac{\psi}{200}\right)^{-1/2} \left(\frac{\rho_{\rm d}}{3}\right)^{1/2}
\end{equation}

\begin{equation}
\dot{M}_{\rm gas} = \psi \dot{M}_{\rm dust},
\end{equation}

where $\psi$ is the gas-to-dust ratio ($\psi = 2891$, assuming [Fe/H]
$= -1.16$ and $\psi_\odot = 200$), $L$ is the stellar luminosity
determined from the {\sc marcs} best fit, and $\rho_{\rm d}$ is the bulk
grain density. For silicates, we assume $\rho_{\rm d} = 3~{\rm
g~cm}^{-3}$, and for AMC grains, we assume $\rho_{\rm d} = 2.5~{\rm
g~cm}^{-3}$. Note that a decrease in metallicity to [Fe/H] $= -1.28$
would result in a 15\% decrease in $\dot{M}_{\rm dust}$. A
summary of input model parameters and output mass-loss rates is
presented in Table~\ref{tab:dusty}.

%%%%%%%%%%%%%%%%%%%%%%%%%%%%%%%%%%%%%%%%%%%%%%%%%%%%%%%%%%%%%%%
% Fig 9: 
%%%%%%%%%%%%%%%%%%%%%%%%%%%%%%%%%%%%%%%%%%%%%%%%%%%%%%%%%%%%%%%

%%%%%%%%%%%%%%%%%%%%%%%%%%%%%%%%%%%%%%%%%%%%%%%%%%%%%%%%%%%%%%%
% Table 3:
\begin{deluxetable}{lcccl}
\tablewidth{0pc}
\tabletypesize{\footnotesize}
\tablecolumns{5} 
\tablecaption{Mass-loss Rates\label{tab:dusty}}

\tablehead{\colhead{Source}&\colhead{$T_{\rm inner}$}&\colhead{Dust}&\colhead{$\dot{M}_{\rm dust}$}&\colhead{$\dot{M}_{\rm gas}$}\\\colhead{ID}&\colhead{(K)}&\colhead{Type}&\colhead{($M_\odot~\rm{yr}^{-1}$)}&\colhead{($M_\odot~\rm{yr}^{-1}$)}}
\startdata

 s02& 500& AMC& $5.9\times 10^{-10}$& $1.7\times 10^{-6}$\\
 s03& 1200& silicate& $2.5\times 10^{-10}$& $7.1\times 10^{-7}$\\
 s04& 1200& silicate& $2.1\times 10^{-10}$& $6.1\times 10^{-7}$\\
 s05\tablenotemark{a}& 1000& silicate& $3.2\times 10^{-10}$& $9.3\times 10^{-7}$\\
 s06& 600& AMC& $7.0\times 10^{-10}$& $2.0\times 10^{-6}$\\
 s07& 800& silicate& $4.6\times 10^{-10}$& $1.3\times 10^{-6}$\\
 s09\tablenotemark{b}& 1200& silicate& $2.6\times 10^{-10}$& $7.4\times 10^{-7}$\\

s10& 1200& silicate& $1.7\times 10^{-10}$& $4.8\times 10^{-7}$
\enddata

\tablecomments{\ Sources s02 and s06 are fit equally well to a {\sc
dusty} model composed of a combination of silicates and AMC dust,
resulting in a nearly identical mass-loss rate to that derived from
the AMC models. See Section~\ref{sec:uncertainties} for a description
of the uncertainties in the mass-loss rates quoted here.}

\tablenotetext{a}{\ The silicate model fits well to the {\it Spitzer}
data for s05. {\it AKARI} data for s05 is better fit by an AMC model with
$T_{\rm inner} = 800$~K, $\dot{M}_{\rm dust} = 5.6\times 10^{-10}
~M_\odot~\rm{yr}^{-1}$, and $\dot{M}_{\rm gas} = 1.6\times
10^{-6}~M_\odot~\rm{yr}^{-1}$.}

\tablenotetext{b}{\ Source s09 is fit equally well to a model with AMC dust
and $T_{\rm inner} = 800$~K.  The resulting mass-loss rates are a
factor of 1.7 less than the rates derived from the silicate
model.}

\end{deluxetable}
%%%%%%%%%%%%%%%%%%%%%%%%%%%%%%%%%%%%%%%%%%%%%%%%%%%%%%%%%%%%%%%

The moderately dusty stars are all fit well by silicate dust. In
addition, source s09 is fit equally well by AMC dust due to a lack of
reliable {\it AKARI} data. The mass-loss rate for s09 from the
best-fit AMC model is a factor of 1.7 less than the rate from the
best-fit silicate model. Together, the four moderately dusty cluster
members lose $\dot{M}_{\rm dust} = 8.3(\pm 0.6) \times 10^{-10}
M_\odot~{\rm yr^{-1}}$, where the errors come solely from the dust
composition used for source s09. This corresponds to $\dot{M}_{\rm
gas} = 2.4(\pm 0.2) \times 10^{-6} M_\odot~{\rm yr^{-1}}$.

It is not surprising to find silicate dust in the outflows of these
four moderate stars. \citet{hofner07} and \citet{hofner08} suggest
that winds could be driven by large ($\sim 1$~\micron{}) Fe-free
silicate grains, or that silicates can easily form in winds that are
driven by carbon grains.

%%%%%%%%%%%%%%%%%%%%%%%%%%%%%%%%%%%%%%%%%%%%%%%%%%%%%%%%%%%%%%%
% Fig 10: 
%%%%%%%%%%%%%%%%%%%%%%%%%%%%%%%%%%%%%%%%%%%%%%%%%%%%%%%%%%%%%%%

\subsubsection{The most extreme dust producers}
\label{sec:extreme}

\begin{figure} %sed_10885.ps
\epsscale{1.2} \plotone{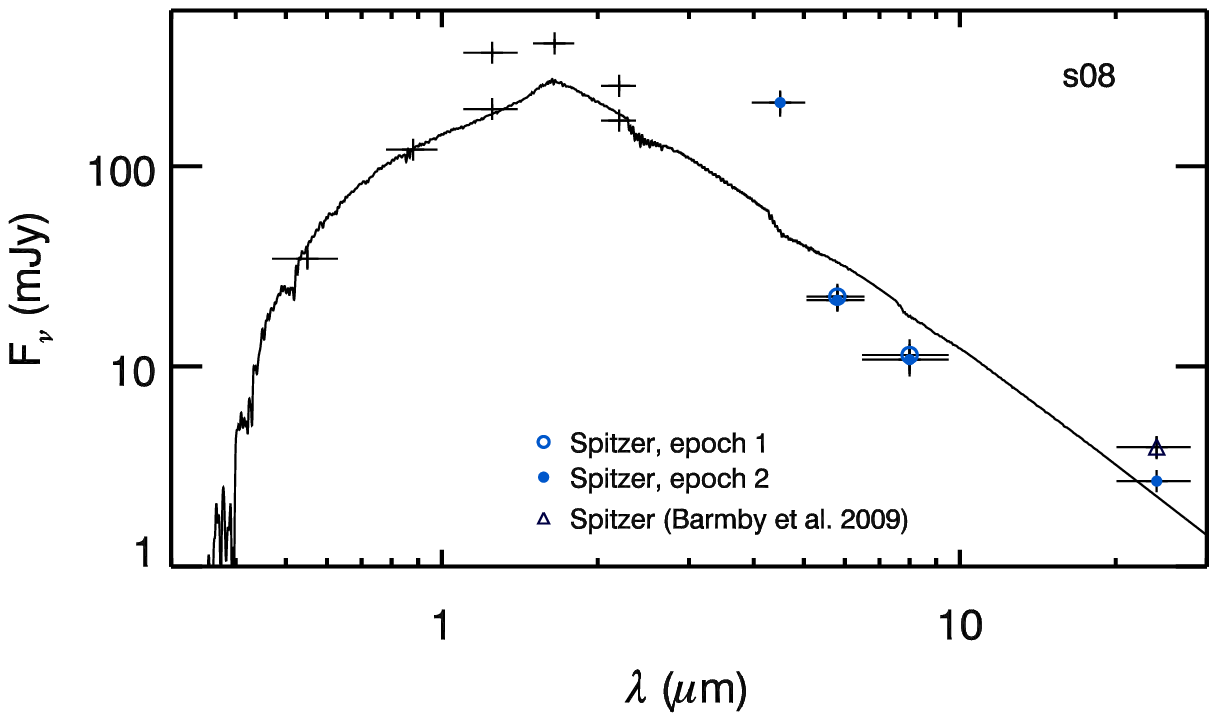} \figcaption{SED for source s08.
We did not attempt to fit a {\sc dusty} model to this SED due to
inconsistent {\it Spitzer} and optical photometry and a lack of {\it AKARI}
photometry. This source may be affected by stellar blending or it may
be extremely variable.\label{fig:10885} }
\end{figure}

Five sources are classified as strong mass-losing stars (s02, s05,
s06, s07, and s08). We do not attempt to fit a {\sc dusty} model to
source s08 due to inconsistent {\it Spitzer} and optical photometry
and a lack of {\it AKARI} photometry (Fig.~\ref{fig:10885}). The remaining
four stars do appear moderately variable in the IR between {\it
Spitzer} and {\it AKARI} epochs (Fig.~\ref{fig:seds_strong}), with s05
showing the largest amplitude followed by s02 and s06, which are the
known variable stars V2 and V16, respectively. The total dust-traced
mass-loss rate from the four extreme stars is $\dot{M}_{\rm gas} =
6.2(\pm 0.5) \times 10^{-6} M_\odot~{\rm yr^{-1}}$, with the
uncertainty depending on the dust compositions chosen. This is more
than twice the contribution of the four more moderate stars.

%%%%%%%%%%%%%%%%%%%%%%%%%%%%%%%%%%%%%%%%%%%%%%%%%%%%%%%%%%%%%%%
% Fig 11: 
%%%%%%%%%%%%%%%%%%%%%%%%%%%%%%%%%%%%%%%%%%%%%%%%%%%%%%%%%%%%%%%

The dust composition is difficult to estimate for the four extreme
sources. Source s05 is particularly odd.  The {\it AKARI} and {\it
Spitzer} data appear to show two different dust compositions. AMC dust
fits best to the {\it AKARI} data, but is simultaneously ruled out by
the {\it Spitzer} data due to a lack of excess between 3.6 and
8~\micron{}. \citet{mcdonald07} found that this source is actually two
stars (o05a/o05b) with similar temperatures and luminosities ($T
\approx 4100$~K, $L_{o05a} = 2103 \pm 189~L_\odot, L_{o05b} = 2392 \pm
213~L_\odot$), which are totally unresolved in both {\it Spitzer} and
{\it AKARI} data.  Both stars appear oxygen-rich and show blue-shifted
H$\alpha$ absorption cores that indicate an outflow of material. It is
unclear how their small separation (0.2\arcsec{}) has affected the IR
photometry. It is possible that the low resolution of the {\it AKARI}
images has resulted in blending of sources in addition to o05a and
o05b, causing inflated {\it AKARI} fluxes. Another possibility is that
one or both of these sources is variable, producing different types of
dust in varying proportions during different phases of the pulsation
cycle. A third possibility is that neither star has a real IR excess,
and the apparent excess is caused by inaccurate flux extraction due to
stellar blending in this crowded region of the cluster. However, this
third possibility seems unlikely since s05 also shows a strong
24~\micron{} excess (Fig.~\ref{fig:seds_strong}), suggesting that at
least one of the stars is producing dust.  Nevertheless, the H$\alpha$
line clearly shows that both stars (o05a/o05b) are losing mass,
although it may not be as clear that the mass loss in either star is
accompanied by dust production.

For source s07, {\it AKARI} photometry is available only at 24~\micron{},
and this flux is more than twice the flux measured with MIPS at
24~\micron{}. If the {\it AKARI} 24~\micron{} flux is accurate, it can only
be explained by an extremely large quantity of circumstellar
dust. This, along with the source's location near several other bright
24~\micron{} sources, suggests that the {\it AKARI} 24~\micron{} flux is
overestimated.  We therefore exclude the {\it AKARI} 24~\micron{} point from
the {\sc dusty} fit and find that silicate dust provides the best fit
due to the lack of strong excess in the IRAC data.

The two remaining extreme sources, s02 (V2) and s06 (V16/o01), are the
most extreme stars in the cluster and exhibit the strongest mass-loss
rates. Neither star is fit well by silicates alone, but both are well
fit by AMC models. This is surprising, given that the optical spectrum
of s06 indicates C/O $<$ 1 \citep{mcdonald07}. \citet{smith99} did not
measure the carbon abundance in star s02, so its nature as carbon or
oxygen-rich remains unknown. Carbon stars {\it are} known to reside in
GCs, notably in $\omega$\,Cen \citep{vanloon07}. Models that include a
combination of 50\% AMC dust and 50\% silicate dust also fit well to
both s06 and s02, but we note that to fit this model to s02, we must
exclude the {\it AKARI} photometry, which shows no silicate features.

A mid-IR spectrum of s02 was obtained by Y.~Ita et al. (in
preparation) using {\it AKARI} IRC slit-less
spectroscopy. Pre-analysis of the spectrum suggests a complete lack of
silicates and the possible presence CO and water absorption near 3 and
5~\micron{}, which may explain the underluminous {\it AKARI}
photometry near these wavelengths. While we cannot definitively
declare a dust type in either s02 or s06, AMC dust must play some role
if we are to reconcile the strong photometric excess in the IRAC bands
and the excess at 24~\micron{}.

\subsubsection{Uncertainties in the Mass-Loss Rates}
\label{sec:uncertainties}

The uncertainties quoted for the total mass-loss rates of the
candidate mass-losing stars reflect a range of values that depends on
the dust composition chosen.  The uncertainties in the mass-loss rates
for individual stars is not reflected in this value.  In $\omega$\,Cen,
uncertainties of 52 -- 80\% were derived for the mass-loss rates of
individual stars, based on internal errors from {\sc dusty} and on
several assumptions.  These assumptions include (1) pulsation causes
only negligible variations in the mid-IR, (2) the gas-to-dust ratio
scales with metallicity, (3) the wind velocity scales as $v_{\rm wind}
\propto L^{1/4} (\psi)^{-1/2}$, and (4) dust-to-gas coupling is
efficient. See \citet{mcdonald09} for a more detailed discussion of
mass-loss rate uncertainties.

We note that the prescription we use to estimate the mass-loss rates
assumes that the wind speed scales as $v_{\rm wind} \propto L^{1/4}
(\psi \rho_{\rm d})^{-1/2}$, resulting in velocities ranging from
$v_{\rm wind} = 0.5 - 1.3~{\rm km~s^{-1}}$ for mass-loss candidates in
NGC~362. These small velocities provide the largest uncertainty to our
mass-loss rates, as they are on the order of the turbulence in the
wind \citep[$\sim$$1 - 2~{\rm km~s^{-1}}$;][]{schoier04}.  It is not
clear whether very slow winds can be sustained, but it is possible
that wind speeds can be increased to $\lesssim$$10~{\rm km~s^{-1}}$ due to
the effects of rotation and pulsation in the star. Moreover,
\citet{mcdonald07} estimated that the wind velocities in several
NGC~362 stars range from $\approx$$6 - 17~{\rm km~s^{-1}}$. If these
velocities are good approximations of the wind speed in the
dust-producing zone, then the mass-loss rates of each star will
increase by up to a factor 6.

\section{Discussion}
\label{sec:disc}

\subsection{Dust and Mass Loss in V2 and V16}
\label{sec:v2v16}

Star s06 (V16) appears to show carbon dust despite its nature as an
oxygen-rich star.  Star s02 (V2) is similar, although its nature as
carbon- or oxygen-rich is unknown. These stars are akin to V42 in
$\omega$\,Cen \citep[cf.][]{mcdonald09}, an oxygen-rich star which
once may have shown a strong 10-\micron{} silicate dust feature, but
more recently appears to show predominantly amorphous carbon dust
despite its oxygen-rich atmosphere. While $\omega$\,Cen V42 and
NGC~362 V16 both appear to be currently dominated by carbonaceous
dust, they may still contain some (possibly variable) amounts of
silicates. 47\,Tuc is home to two dusty stars (V1 and V18) that show
variable silicate emission \citep{vanloon06a,
lebzelter06}. \citet{hofner07} suggest a possible scenario where
non-local thermodynamic equilibrium conditions allow the periodic
production of different grain compositions at different epochs in the
pulsation cycle. It may also be the case that grain size varies enough
that large grains periodically suppress the silicate features.

\subsection{Global Mass Loss}
\label{sec:globalmlr}

The cluster stars that exhibit IR excesses are currently returning
$\dot{M}_{\rm gas} = 8.6^{+5.6}_{-3.4} \times 10^{-6} M_\odot~{\rm
yr^{-1}}$ of gas to the ICM, using the same fractional errors used in
\citet{mcdonald09}.  To find the total cluster mass loss (dusty and
dustless), we estimate the mass-loss rates for non-dusty stars using
the method described in \citet{schroder05}, which is based on the
stellar temperature and luminosity. We find that
chromospherically-driven (dustless) mass loss is responsible for $2
\times 10^{-6} M_\odot~{\rm yr^{-1}}$, giving a total cluster mass
loss of $\dot{M}_{\rm total} \approx 1.1^{+0.7}_{-0.4} \times 10^{-5}
M_\odot~{\rm yr^{-1}}$. Dust production therefore accompanies 80\% of
the current mass loss in NGC~362.

The global mass-loss rate for NGC~362 is virtually identical to that
of $\omega$\,Cen \citep[$\approx 1.2 \times 10^{-5}~M_\odot~{\rm
yr^{-1}}$;][]{mcdonald09}, despite NGC~362 being an order of magnitude
less massive than the latter. 47\,Tuc, which is similar in mass to
$\omega$\,Cen, appears to have a similar global mass-loss rate, given
that its four long-period variables have mass-loss rates ranging from
$10^{-7}$ to $10^{-6} M_\odot~{\rm yr^{-1}}$ \citep{origlia07}. In all
three clusters, only a handful of stars are dominating the global
mass-loss rate. This supports the likelihood that global dust
production and dust-associated mass loss in clusters are stochastic in
nature, driven by episodic dust production in the individual stars
(possibly due to varying stellar conditions at different pulsation
phases) and the rarity of the stars themselves.  $\omega$\,Cen may
currently be experiencing a relatively quiescent period of dust
production, given that only two stars (V6 and V42) are currently
forming significant amounts of dust.  M15, on the other hand, appears
to have recently experienced a period of strong dust production,
assuming that the ICM dust present there was formed in stellar winds
rather than by a stellar collision \citep{rasio91} or one or more
diffusing planetary nebulae. NGC~362 has a moderate total mass ($3.78
\times 10^5~M_\odot$) and up to four stars producing significant
amounts of dust. NGC~362 may therefore be representative of a more
intermediate phase of dust production in the typical GC.

Any effects that metallicity has on dust production appear to be
overshadowed by the apparent episodic nature of dust production. Stars
with strong IR excess are discovered in clusters with metallicities
ranging from ${\rm [Fe/H]} = -2.4$ for M15 \citep[e.g.,][]{boyer06} to
${\rm [Fe/H]} = -0.7$ for 47\,Tuc \citep[][]{vanloon06a}, with no
clear trend favoring higher metallicities. 

Dust production is dominated by the most luminous stars in the
cluster. However, several dust{\it less} stars have temperatures and
luminosities similar to the dust-producing stars. The H$\alpha$
profiles of at least a handful of these dustless stars suggest that
they are losing mass without any associated dust production.  The only
observed difference known to exist between the dusty and non-dusty
stars is a lack of a red H$\alpha$ line emission wing in the non-dusty
stars, perhaps suggesting that material is restricted to smaller
stellar radii than in the dusty stars \citep[cf.][]{mcdonald07}.

\subsection{The Intracluster Medium}
\label{sec:icm}

If we make the (possibly na\"{i}ve) assumption that the rate of gas
return has remained relatively constant since NGC~362 last plunged
through the Galactic plane ($3 \times 10^7$ yr ago), then we might
expect 210 -- 540 $M_\odot$ of gas and 0.05 -- 0.15 $M_\odot$ of dust
to have gathered in the ICM, especially considering that the cluster
escape velocity is much larger than typical wind speeds derived with
{\sc dusty} ($v_{\rm esc,0} = 46.7~{\rm km~s^{-1}}$, $v_{\rm wind} =
0.5 - 1.3~{\rm km~s^{-1}}$, assuming $v_{\rm wind} \propto L^{1/4}
(\psi \rho_{\rm d})^{-1/2}$). Rotation and magnetic activity and
pulsation shocks can increase wind velocity by up to an order of
magnitude \citep[cf.][]{bowen88}, but this increase is still not
sufficient for material to escape the cluster. Within the half-light
radius, upper limits of only $<6.3 \times 10^{-5} M_\odot$ of dust
\citep{barmby09} and $<1.8~M_\odot$ of ionized hydrogen
\citep{hesser77} have been determined in the ICM of NGC~362.

Assuming that $\lesssim$5\% of the ICM gas is ionized, as in 47\,Tuc
\citep{smith90,freire01}, we find that there is $<40~M_\odot$ of
neutral gas in the ICM of NGC~362.  Based on the total (dust $+$
dustless) cluster mass-loss rate ($1.1 \times 10^{-5}~M_\odot~{\rm
yr^{-1}}$), we can expect neutral gas to be cleared from the cluster
on timescales $<3.6 \times 10^6$~yr, more than an order of magnitude
shorter than the time since the last Galactic plane-crossing.  The
dust provides an even shorter timescale of ICM removal or destruction,
namely $<2 \times 10^4$~yr, given a dust mass-loss rate of $3 \times
10^{-9}~M_\odot~{\rm yr^{-1}}$ and a dust upper limit of $<6.3 \times
10^{-5} M_\odot$. To escape the half-mass radius (0.81\arcmin{})
within 20\,000~yr, the dust would have to travel $>80~{\rm
km~s^{-1}}$.  This velocity is much higher than the likely wind
velocities \citep[$v_{\rm wind} \lesssim 10~{\rm
km~s^{-1}}$,][]{mcdonald07,meszaros09}, requiring that (a) either the
dust is dissociated on short timescales within the cluster or that (b)
dust is rapidly accelerated out of the cluster.

Based on the space velocity of NGC~362, \citep[140~km~s$^{-1}$,
cf.][]{odenkirchen97}, ram-pressure from hot Galactic Halo gas could
clear the cluster ICM in $\approx$$2.8 \times 10^5$~yr. A stellar
collision has enough kinetic energy to clear the ICM, and such
collisions occur on timescales of $\approx$$2.5 \times 10^6$~yr in
NGC~362 \citep[cf.][]{barmby09}. If ICM {\it dust} were destroyed
within the cluster on the required shorter timescales, then
ram-pressure stripping or a stellar collision may be responsible for
the lack of gas and the byproducts of dust destruction in NGC~362.

Although mass loss estimates predict the presence of
$\sim$$10^2~M_\odot$ of ICM material in most GCs, nothing approaching
this amount of mass has ever been observed. $\omega$\,Cen has an upper
limit of $<10^{-4}~M_\odot$ of dust \citep{boyer08} and $<2.8~M_\odot$
of neutral gas \citep{smith90}. \citet{barmby09} finds ICM dust mass
upper limits of $<10^{-4}~M_\odot$ in eight GCs, and
\citet{matsunaga08} finds upper limits of $<10^{-3}~M_\odot$ of dust
in 12 GCs. \citet{vanloon09} determined 3\,$\sigma$ upper limits
between 6 and 51~$M_\odot$ of neutral hydrogen within the tidal
volumes of four GCs.  Much of the ICM may be ionized, as demonstrated
in 47\,Tuc, where up to $0.1~M_\odot$ of ionized gas was discovered
\citep{freire01}.  M15 shows the strongest evidence for an ICM, with
firm detections of $9 \times 10^{-4}~M_\odot$ of dust \citep{boyer06}
and $0.3~M_\odot$ of neutral hydrogen \citep{vanloon06b}. The
combination of episodic dust production and a stochastic ICM removal
process (such as a stellar collision) could explain the lack of dusty
ICM material in GCs and its unusual presence in M15. If ram-pressure
stripping from hot Halo gas is the dominant removal mechanism, the
intense crowding in M15 \citep[log($r_{\rm tidal}/r_{\rm core})=2.5$,
compared to 1.9 for NGC~362;][]{harris96} could cause the stripping of
material in the center of the cluster to be less efficient, with
bow-shocks around outer stars effectively shielding the cluster's ICM.

It has become clear that low-mass, low-metallicity stars successfully
form dust and lose mass. Therefore, some mechanism must remove gas and
dust from GCs.  This material may ultimately reside in the Galactic
Halo, fall back onto cluster stars, or find its way back into the
Milky Way disk \citep[e.g.,][]{evans03,boyer06,vanloon06b,vanloon09}.
In any case, low-mass, low-metallicity stars like those in globular
clusters must collectively and continuously contribute a large amount
of recycled and enriched material to the ISM, helping to drive galaxy
evolution.

\section{SUMMARY OF RESULTS AND CONCLUSIONS}
\label{sec:summary}

We present an analysis of dusty mass loss in the Galactic globular
cluster NGC~362 carried out with data serendipitously obtained during
{\it Spitzer} SAGE-SMC observations of the Small Magellanic Cloud.
Spectral energy distribution modeling of all cluster stars provided
stellar parameters, allowing the construction of a physical
Hertzsprung-Russell Diagram. The Red Giant Branch (RGB) is slightly
cooler than the Padova isochrone, indicating significant mass loss on
the RGB.

Significant infrared (IR) excess exists only at and above the tip of
the RGB. The four brightest stars at 24~\micron{} exhibit the
strongest IR excess. Four additional stars near the tip of the RGB
show moderate IR excess. Modeling with the {\sc dusty} code indicates
that these eight stars contribute $8.6^{+5.6}_{-3.4} \times 10^{-6}
M_\odot~{\rm yr^{-1}}$ of gas and $3.0^{+2.0}_{-1.2} \times 10^{-9}
M_\odot~{\rm yr^{-1}}$ of dust to the intracluster medium. The two
most extreme stars, variables V2 and V16, provide up to 45\% of the
total (dust accompanied) mass loss.

The eight dustiest stars show evidence of silicate dust, but,
surprisingly, three of the four more extreme stars require some amount
of amorphous carbon dust to explain their mid-IR excesses. Strong mass
loss therefore appears to correlate with a larger contribution from
carbon dust, which may suggest that non-equilibrium conditions are
common in such stars.

\acknowledgements We thank Jay Anderson for sharing the Hubble ACS
image and Yoshifusa Ita for sharing {\it AKARI} photometry and the
mid-IR spectrum of star s02. We also thank the referee for his or her
helpful comments. This work was supported by {\it Spitzer} via JPL
contracts 1309827 and 1340964.

%%%%%%%%%%%%%%%%%%%%%%%%%%%%%%%%%%%%%%%%%%%%%%%%%%%%%%%%%%%%%%%
%%%% Bibliography
%%%%%%%%%%%%%%%%%%%%%%%%%%%%%%%%%%%%%%%%%%%%%%%%%%%%%%%%%%%%%%%%

\clearpage

\end{document}